\documentclass[useAMS,usenatbib,usegraphicx]{mn2e}
\usepackage{amssymb,amsmath}
\title[The Grinding Machine] {Intergalactic stellar populations in intermediate redshift clusters. }

\author[J. Melnick et al.]  { J.~Melnick,$^1$ E. Giraud,$^2$ I.~Toledo,$^3$ F.~Selman,$^1$ H. Quintana $^3$ \\
   $^1$European Southern Observatory, Alonso de C\'ordova 3107, Santiago, Chile\\
   $^2$Laboratoire Physique Th\'eorique et Astroparticules (LPTA), Universit\'e Montpellier 2 - CNRS/IN2P3, \\ Place E. Bataillon, 34095 Montpellier, France\\
   $^3$Departamento de Astronom\'ia y Astrof\'isica, Pontificia Universidad Cat\'olica de Chile, Casilla 306, Santiago, Chile
  }

\begin{document}

\date{}

\pagerange{\pageref{firstpage}--\pageref{lastpage}} \pubyear{2012}

\maketitle

\label{firstpage}

\begin{abstract}
A substantial fraction of the total {\it stellar} mass in rich clusters of galaxies resides in a diffuse intergalactic component usually referred to as the Intra-Cluster Light (ICL). Theoretical models indicate that these intergalactic stars originate mostly from the tidal interaction of the cluster galaxies during the assembly history of the cluster, and that a significant fraction of these stars could have formed in-situ from the late infall of cold metal-poor gas clouds onto the cluster. However, these models also over-predict the fraction of stellar mass in the ICL by a substantial margin, something that is still not well understood. The models also make predictions about the age distribution of the ICL stars, which may provide additional observational constraints. Here we present population synthesis models for the ICL of an intermediate redshift (z=0.29) X-ray cluster that we have extensively studied in previous papers. The advantage of observing intermediate redshift clusters rather than nearby ones is that the former fit the field of view of multi-object spectrographs in 8m telescopes and therefore permit to encompass most of the ICL with only a few well placed slits.

In this paper we show that by stacking spectra at different locations within the ICL it is possible to reach sufficiently high S/N ratios to fit population synthesis models and derive meaningful results. The models provide ages and metallicities for the dominant populations at several different locations within the ICL and the BCG halo, as well as measures of the kinematics of the stars as a function of distance from the BCG. We thus  find that the ICL in our cluster is dominated by old metal rich stars, at odds with what has been found in nearby clusters where the stars that dominate the ICL are old and metal poor. While we see a weak evidence of a young, metal poor, component, if real, these young stars would amount to less than 1\% of the total ICL mass, much less than the up to 30\% predicted by the models.

We propose that the very metal rich (i.e. $2.5\times$ solar) stars in the ICL of our cluster, which comprise $\sim40$\% of the total mass, originate mostly from the central dumb-bell galaxy, while the remaining solar and metal poor stars come from spiral, post-starburst (E+A), and metal poor dwarf galaxies. About 16\% of the ICL stars are old and metal poor.
 
\end{abstract}


\section{Motivation}
\label{SECintro}

In the 1930's Fritz Zwicky had a series of profound astronomical insights. One of them was that intergalactic space could not be just empty but had to be populated by stars, dust, and gas. This  motivated him to look for intergalactic stars and supernovae in a number of rich clusters using first the 18-inch and later the 48-inch Schmidt telescopes on Palomar mountain  \citep{Zwicky1952}.  These searches led him to discover bridges of light between a triplet of adjacent galaxies in the Coma cluster that he attributed to stars ``escaping from the triple system entirely" and making their way into intergalactic space. Thus, Zwicky not only pioneered the field of diffuse intergalactic light in clusters of galaxies, but also correctly guessed  the origin of that light.

Still, the field remained dormant for almost 20 years until it was revived in the 1970's when it became understood that the missing mass in rich clusters of galaxies (which incidentally was also first pointed out by Zwicky in 1933), had to be diffuse and not concentrated in the cluster galaxies themselves ( \cite{deVaucouleurs1970}; \cite{Melnick1977}; \cite{Thuan1977}). These early photoelectric and photographic investigations indicated that the true intergalactic light did not exceed $\sim25\%$ of the total stellar light of the Coma cluster and, therefore, that the material that binds the cluster had a very high mass-to-light ratio. It was also understood in the 70's that the strong X-ray emission that emanated from rich galaxy clusters was produced by thermal Brehmstrahlung  and, therefore, that  the gravitational potential wells of  rich clusters were filled with large amounts of hot gas. But even the mass of hot gas was not enough to solve the missing mass problem, and the diffuse stellar intergalactic medium was again forgotten for the next 20 years.  

A third revival has taken place in the last decade through a combination of realistic numerical simulations and powerful new observational techniques. Numerical simulations have shown that, as Zwicky predicated 50 years ago, tidal interactions during the merging history of clusters `release' large numbers of stars that make their way into the intergalactic space carrying with them the memory of the evolutionary history of clusters. In principle, therefore, the diffuse stellar component - which we now call the intracluster light or ICL - contains a fossil record  of the star formation history of clusters of galaxies.   In practice, however,  the comparison between observations and cosmological simulations yields surprisingly weak constrains. This is partly due to the fact that in both simulations and observations it is exceedingly difficult to disentangle bona-fide free-floating stars from the halos of the brightest cluster galaxies (BCG), and partly to the fact that at a given mass there is a manifold of possible merging histories (see e.g. \cite{Puchwein2010} and references therein).  Comparison of theory and observations, therefore, require representative samples of clusters and thus poses formidable observational (and non trivial numerical) challenges.  

The current state-of-the-art results indicate, for example,  that while AGN feedback reduces the discrepancy between models and observations, the ICL fractions observed in a reasonably representative sample of clusters are still significantly lower than the fractions predicted by the models (\cite{Puchwein2010} and references therein). A surprising prediction from these models (with or without AGN feedback) is that as much as 30\% of the intergalactic stars are actually born in `cold' intergalactic gas clouds stripped off galaxies by tidal interactions. Intergalactic star formation has been verified observationally in the intergalactic medium (see e.g. \cite{Sun2010} and \cite{Boquien2010} for recent references), but what is unexpected is that so much of the ICL would come from these events. In fact, the discrepancy between ICL fractions in models and observations would basically disappear if  these stars did not exist.  Therefore, studies of the stellar populations of the ICL promise to  provide useful new constrains to the theoretical models. Until recently, however, it has only been possible to measure the colors of the ICL, which is a notoriously difficult observational task. These colors appear to be systematically bluer than those of the BCG halos, consistent with a scenario where much of the ICL originates in smaller galaxies \citep{Rudick2010}. Also, the ACS on HST resolved the stellar populations of the ICL in a field of the Virgo cluster and found that, as expected, the intergalactic stars span a wide range of ages and metallicities, but the dominant population accounting for 70-80\% of the stars is old and metal poor \citep{Williams2007}. A similar result was obtained by \cite{Coccato2012} in the cluster Hydra-I (A1060) using long slit spectroscopy Lick indices. In Hydra-I, basically 100\% of the ICL stars appear to be old and metal poor.

In this paper we present  a spectroscopic analysis of the stellar populations in the ICL of an X-ray cluster at z=0.29, RXJ0054.0-2823,  that we extensively studied photometrically and spectroscopically \citep{Toledo2011}. We find that the majority of the ICL stars in this cluster are old, as is the case for the Virgo cluster, but the metallicities are significantly higher indicating a different origin that may be related to the fact that our cluster has three interacting giant elliptical galaxies in the center that conform a terrific galaxy {\it grinding machine}. The paper is organized as follows:  In Section~\ref{SECobs} we recall the data we used in the investigation from previous papers;  Section~\ref{SECpop} presents population synthesis models of our stacked spectra of the ICL; and  Section~\ref{SECres}  presents the comparison of our observations with previous observations and state-of-the-art cosmological numerical simulations. the results and presents the conclusions of this work.   Finally, in Section \ref{SECconcl} we summarize our results and outline future investigations.

\section{The data}
\label{SECobs}

A detailed description of the data-sets used in the present investigation has been presented in three previous papers:  \cite{Giraud2010,Giraud2011}, and \cite{Toledo2011}; henceforth Paper~I, to which we refer for comprehensive discussions of the observing and data analysis methodologies. It is useful to reproduce here the relevant figures from Paper~I  to set the data in the present context, in particular because, as described above, observationally it is extremely difficult, if not impossible, to separate the ICL from the extended halos of the brightest cluster galaxies.

\begin{figure*}
  \includegraphics[height=7.0cm,width=7.5cm]{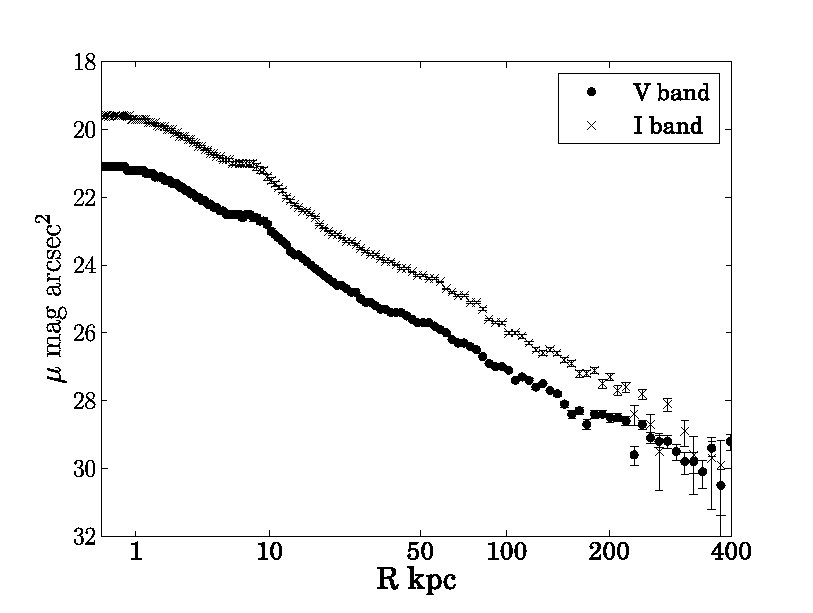}\hspace*{0.1cm}\includegraphics[height=7.0cm, width=7.5cm]{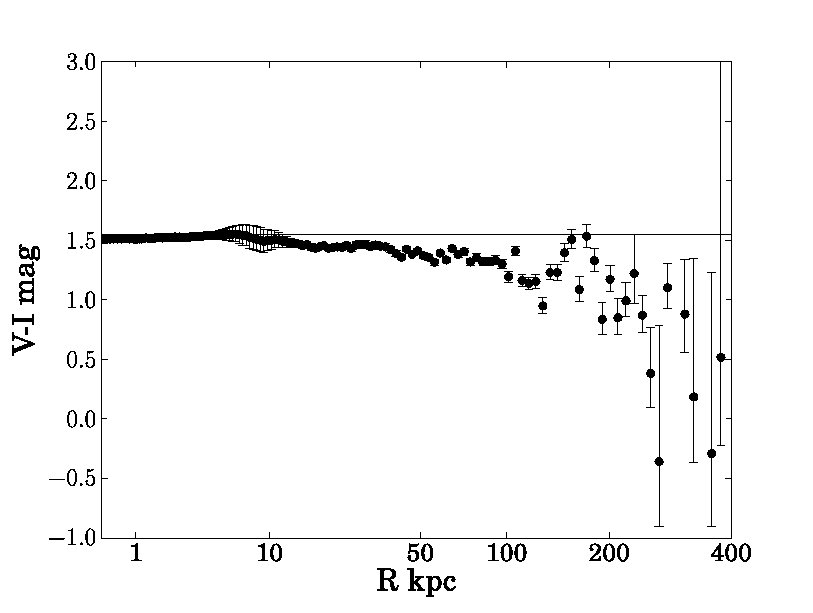}
  \caption{\bf Left \rm Surface brightness profiles in V and I. \bf Right \rm (V-I) color of the BCG+ICL component as a function of radial distance. Special care was taken to match the fits of the two bands to avoid spurious differences due to misalignment and rebinning. Thus, the photometric error-bars provide accurate estimates of the reality of the various features visible in this plot. Both plots are on an $r^{1/4}$ scale.}
  \label{fig:profiles}
\end{figure*}

Following previous investigations (e.g. \cite{Gonzalez2005}; \cite{Gonzalez2007}) we were constrained to use the ICL and the BCG together and to rely on the surface brightness profiles to delineate the most likely boundaries between the two components. These profiles from Paper~I are reproduced in Figure~\ref{fig:profiles}. The photometric error bars, which include residual alignment errors between the V and I images, indicate that we detect the ICL out to at least 400~kpc from the cluster center.   Both the surface brightness and the color profiles exhibit clear features at about $r\sim50$~kpc that we identified in Paper~I as the radius beyond which the diffuse component is dominated by free-floating ICL stars. Thus, for $r<50$kpc we see a mix of BGC and ICL stars, while at $r\geq50$kpc we observe pure ICL light.  We have therefore stacked together our spectra in these regions to reach S/N ratios  suitable for studying their stellar populations, which we discuss in the following Section.  


\section{Population Syntesis}
\label{SECpop}

We fitted simple stellar population (ssp) models to our data using the code {\bf\tt STARLIGHT} described in detail by \cite{Cid2005}.  We used a base of 39 ages ranging from 6~Myr to 11~Gyr and 3 different metallicities Z=0.004, Z=0.02, and Z=0.05, (i.e. metal-poor: 0.2 times solar; solar; and metal-rich: 2.5 times solar) based on the {\bf\tt Galaxev} code \citep{BC03}, the Padova 1994 evolutionary tracks \citep{Padova94}, and the MILES library \citep{MILES}. {\bf\tt STARLIGHT} calculates the best fitting linear combination of  ssp model spectra from this BC94c library described in detail by  \cite{GonzaCid2010} in their comprehensive comparison between different stellar libraries on the basis of  synthesizing the integrated spectra of star clusters mostly in the Magellanic Clouds. We used the CCM extinction law  \citep{Cardelli1989} and a \cite{Chabrier} IMF, although these choices do not affect our results in any significant way.  

As a check, we ran a number of models using different stellar libraries (or bases) kindly provided by Roberto Cid-Fernandez and partly discussed by  \cite{GonzaCid2010} in order to test the sensitivity of our results to the choice bases.  The best fitting stellar populations do depend on the choice of stellar bases, but unfortunately these bases do not cover the same wavelength ranges, the same age ranges, nor the same metallicity ranges. Therefore, as discussed by \cite{GonzaCid2010}, it is difficult to choose the library best suited for a particular set of observations.  

The four libraries we tested (CB07c\footnote{The CB07 base is similar to the BC00 base of \cite{GonzaCid2010} but using the MILES library instead of STELIB.}; BC94c; V00c; and V00s, where $s$ stands for Salpeter and $c$ for Chabrier IMFs) yield very similar fits to the data (as indicated by the $\chi^2$ parameter), but quantitatively different stellar populations. However, all libraries yield the same kinematics, and as mentioned above, the synthesis results are insensitive to the choice of IMF.  Thus, our choice of libraries was mostly dictated by the fact that  BC94 (s or c) span a wider range of ages and metallicities than the V00 models. (In hindsight, the fact that the BC libraries also span a much broader range in wavelength makes them more suitable for redshifted objects.) Therefore, while the diffuse light in our cluster displays a rich mix of stellar populations with any of these libraries, the populations differ in detail for the different libraries, so one must be careful in interpreting the results.
 
\begin{figure*}
  \includegraphics[height=11.5cm,width=12.0cm]{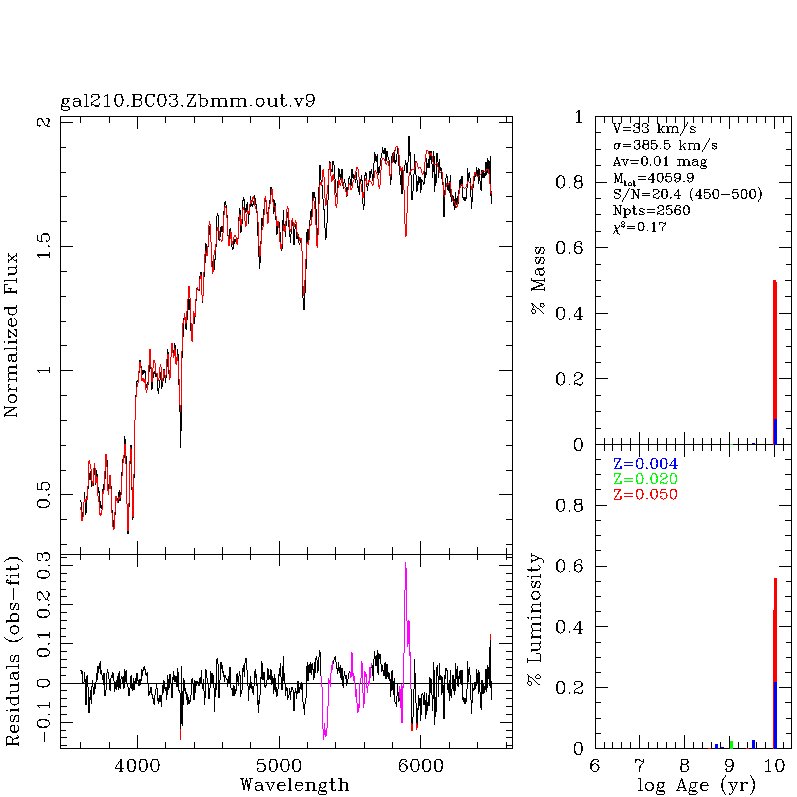}
    \caption{\bf Left panels. \rm The observed spectrum of galaxy \#210 (the fifth ranked elliptical in the cluster) compared to the results of the population synthesis {\bf\tt STARLIGHT} fit in red. The lower panel shows the residuals of the fit. The regions colored in magenta correspond to telluric features that have been masked from the fit. The residuals in red have been automatically excluded by  {\bf\tt STARLIGHT} using a $\kappa-\sigma$ clipping algorithm. \bf Right panels. \rm The right panels show the results of the stellar population fit for three different metallicities as indicated in the Figure. The upper panel shows the fractional contribution of each component by present-day mass, and the lower panel by luminosity. Galaxy \#210 is a typical elliptical where most of the stars are old and metal-rich (red), with a small contribution of old and metal-poor stars (blue).}
   \label{fig:gal210}
\end{figure*}

In order to illustrate the way in which we will present the spectral synthesis results for the ICL, we applied {\bf\tt STARLIGHT} to our high S/N spectrum of galaxy \#210 (cf. Figure~\ref{fig:slits}) which was included in the long slit that we used to observe the ICL along the northern side of the major axis of the brightest cluster galaxy, which turns out to be a dumb-bell system (Paper~I). Figure~\ref{fig:gal210} shows in the left panels the observed (black) and synthetic (red) spectra, normalized to the fluxes at 400-410~nm, and the fit residuals calculated subtracting the model fit from the observed (normalized) spectra. The magenta colors in the residuals panel indicate wavelengths that have been masked from the fit due to contamination from poor subtraction of atmospheric absorption bands; The red colors correspond to data excluded by {\bf\tt STARLIGHT} using a $\kappa-\sigma$ clipping algorithm. Unfortunately, the green OI auroral line at 557.7~nm falls on top of the metallicity-sensitive  G-band at 430.4~nm, which explains why this strong feature has typically high residuals in our spectra.

The right panels of the plot show the population analysis using different colors for the three metallicities. Present-day mass fractions on top, and luminosity fractions on the bottom.  The labels of these figures show the metallicities of the base as well as several parameters resulting from the population fits, notably the velocity dispersion ($\sigma$), and the total visual extinction ($\rm A_V$). The average S/N of the spectrum between 450-500~nm is also given together with an estimate of the goodness of fit measured by the reduced $\chi^2$.


%


Galaxy \#210 is the fifth ranked elliptical galaxy in the cluster with a V magnitude of 19.5, which at the distance of the cluster, $D\sim 1200$ Mpc, makes it an L* galaxy. It's stellar populations are dominated by old metal rich stars ($92\%\pm2\%$ with a small but significant contribution of old metal-poor stars $7.5\pm3\%$, and an insignificant (at the $<1\sigma$ level), residual at intermediate-ages. The extinction is negligible as expected for this high latitude cluster. In fact, the model extinction provides an additional criterion for validating the results, since a high value would be indicative of errors in the flux calibration of the spectra.

The velocity dispersion  uncorrected for instrumental broadening is $\sigma_{obs}=385.2$~km/s. The instrumental broadening measured by fitting a Gaussian profile to the 557.7~nm night-sky feature is 15.5\AA\ FWHM, which corresponds to a velocity dispersion $\sigma_{inst} = 354$ km/s. Thus, the intrinsic velocity dispersion of Gal\#210 is 152~km/s, typical of elliptical galaxies of similar luminosity (e.g. \cite{Chun}).

\subsection{Modeling errors}
\label{SECerrors}

For a variety of reasons obtaining reliable estimates of the statistical uncertainties in the stellar populations derived from ssp model fits is a notoriously difficult task (see e.g.  \cite{GonzaCid2010} for a recent discussion). For spectra with accurate flux calibrations and reasonable S/N ratios (say S/N$>15$), the random errors are small compared to the systematic errors associated with the choice of stellar libraries, the wavelength range of the fit, the masking of certain features, etc.  Thus, understandably albeit unfortunately, {\bf\tt STARLIGHT} does not provide formal errors in the fitted parameters.

In order to estimate the intrinsic errors in our populations, therefore, we ran each model 10 times to compute the dispersion due to numerical errors, and we repeated the operation masking some wavelength regions where the spectra are particularly noisy, or the residuals particularly large. Naturally, the errors depend on the quality of the spectra ranging from $<2\%$ for Gal210 to $\sim10\%$ for the S-shaped Arc (ARCsum) in population ages; but are significantly more stable for the pop. metallicies: $<1\%$ in Gal210 and $<5\%$ in the Arc. The systematics are sensitive to age and metallicity and in general are substantially larger, in particular for the ages. In the Arc, for example, the variance of different masking experiments can reach up to $\sim20\%$ for the ages, but is only $\sim 7\%$ for the metallicities.

{\it A fortiori}, therefore, the errors have an objective random numerical fitting error, and a somewhat subjective systematic component that is estimated by observing the response of the fits to small changes in the input parameters. In the paper we will quote the average values with formal one-$\sigma$ errors, and as relevant, we will comment upon the effects changes in the input parameters, which we will refer to as   {\it systematic variances}.

\subsection{ICL population synthesis}
\label{SECicl}

\begin{figure*}
  \includegraphics[height=10.0cm,width=10.0cm]{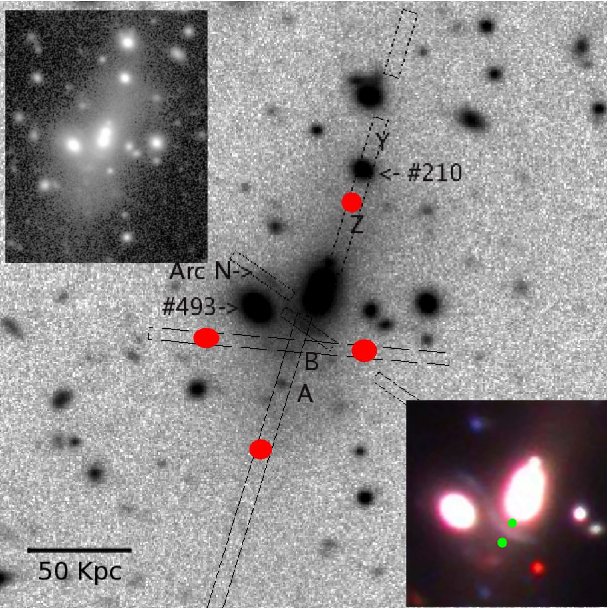}
    \caption{ Approximate positions where the various components of the ICL were measured. For the `pure' ICL we averaged together the spectra in the locations indicated by the red spots. The insert on the top-left of the figure shows the image of the cluster with a stretch optimized to show the ICL. For the inner halo of the BCG we used the spectrum at the position of the green spot located just above the S-shaped arc in the color insert (bottom-right). The outer green spot indicates the mix spectrum between ICL and halo. We also fitted the spectrum at the intersection of the NS and EW slits (clearly visible in the figure marked with a B),  and the spectrum of the sum the two components of the S-shaped arc (arcSUM). The positions of the slitlets used for sky subtraction are also shown in this figure. }
   \label{fig:slits}
\end{figure*}

Figure~\ref{fig:slits} reproduced from Paper~I illustrates the way the long slits were placed on the cluster. Full details of the observations are given in that paper. In order to investigate the properties of the ICL, we integrated the pixels at the locations indicated in Figure~\ref{fig:slits}. The red spots depict the 4 areas that we averaged together to obtain the `pure' ICL spectrum. Thus, the average of these 4 spectra is what we will call ICL-spectrum throughout the rest of the paper.  We also fitted the spectrum of the halo of the BCG (shown by the northern green spot in the inset of Figure~\ref{fig:slits}), and a mixed spectrum of the ICL and the  BCG halo indicated by the southern green spot in the figure. We will call these spectra `inner halo' and `outer halo' respectively. Finally, we used two slitlets to measure the two components of the S-shaped arc and we call the sum of these two spectra {\it arcSUM}.  


\subsection{The BCG Halo and the S-shaped arc}

%


            
Observationally it is very difficult to disentangle the bona-fide intergalactic component (ICL) from the halo of the brightest cluster (dumb-bell) galaxy (BCG; not numbered but clearly visible in Figure~\ref{fig:slits}), which if the cluster has a significant ICL, can be very extended.  Taking advantage of the fact that we used long slits, we are able to sample the diffuse light at various distances from the BCG and look for systematic trends. 

\begin{figure*}
 \hspace*{-0.3cm} \includegraphics[height=8.0cm,width=8.5cm]{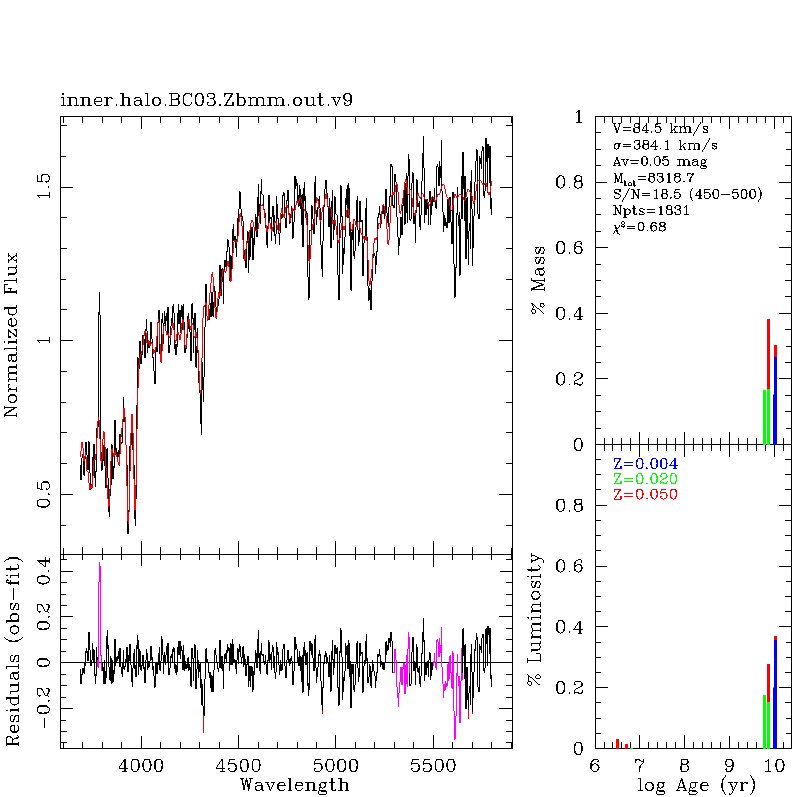}\hspace*{0.5cm}
 \includegraphics[height=8cm,width=8.5cm]{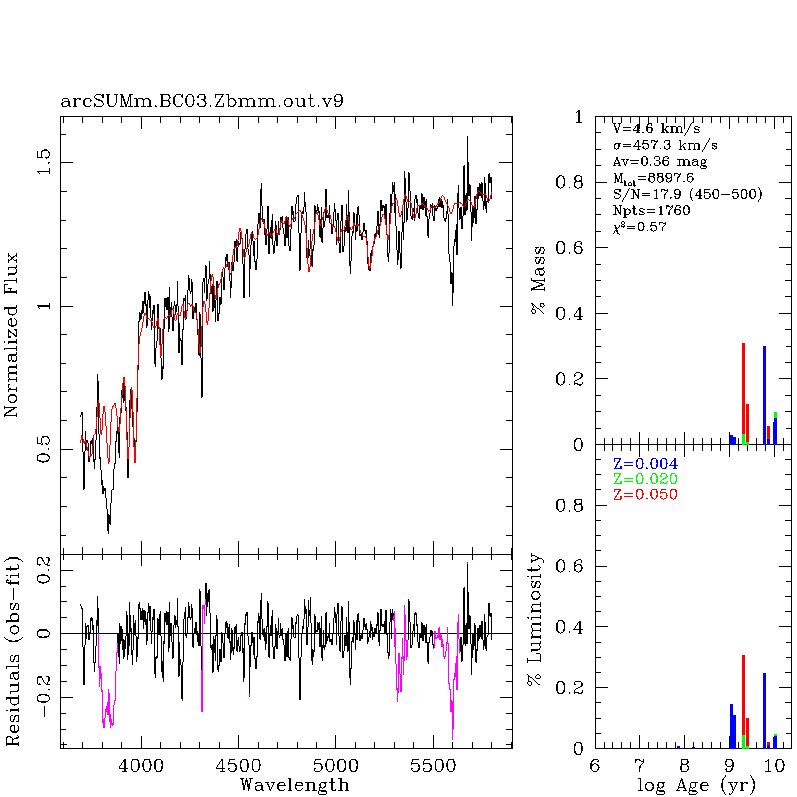}
    \caption{ \bf Left panel: \rm Spectrum and population synthesis fits of the BCG halo at the position indicated by the green spot above the Arc in Figure~\ref{fig:slits}. The strong emission feature at $\sim3800$\AA\  is not real and was masked from the fit.  \bf Right panel:  \rm   Spectrum and fit to the sum of the two arc components (N\&S) that were covered by two different slitlets are shown in Figure~\ref{fig:slits}. The strong CN feature at 380-385~nm could not be fitted by any models and therefore was masked-out to avoid introducing systematic errors. }
   \label{fig:halo}
\end{figure*}

The left panel of Figure~\ref{fig:halo} shows the spectrum and the ssp fit for the inner BCG halo in exactly the same format we used for Gal\#210 above. The position of the inner halo is indicated in the inset of Figure~\ref{fig:slits} by the green spot just above the Arc.  Somewhat surprisingly, while close to 85\% of the stars in the inner halo are old, only about  $20\pm12\%$ are metal-rich while $42\pm1\%$ are old and metal-poor. The reminder have solar metallicity and either old or intermediate ages. The old metal-poor population remains stable for different experiments, but the systematic variances
of the other populations are rather large. The strong old metal-poor component indicates that a substantial fraction of  the BCG halo is made of stars accreted from ancient low-mass galaxies.  

The velocity dispersion of the BCG Halo (150~km/s) is typical of the central velocity dispersions of luminous elliptical galaxies, but seems too large for the outer  halos of these galaxies (e.g. \cite{Carollo1995} ). However, in cD galaxies the velocity dispersions rise towards the edges of the halo \citep{Kormendy1989}. Unfortunately we do not have a complete long-slit spectrum of the BCG to check whether we observe a steep velocity dispersion profile, such as seen in M87 for example, or a rather flat profile like those observed in normal ellipticals  (\cite{Doherty2009} and references therein).  

The velocity dispersions of all components are shown in Table~\ref{tab:kine}, and the radial velocities of the individual components before averaging is given in Paper~I.    
 
%
            


 The right-hand panel of Figure~\ref{fig:halo} shows the spectrum of the S-shaped arc (arcSUM), which appears to be a spiral galaxy in the process of being disrupted by the tidal field of the `grinding machine'.  Spectroscopically the galaxy is a typical E+A or `post-starburst' galaxy. It shows a very complex set of populations, although it should be apparent from the figure that the models are particularly sensitive to the masking of different features in the spectrum. The most stable population is intermediate age ($<7.5\times10^9$~yr) and metal rich, with $42\pm10\%$ of the stars.  This component could come from a galaxy-wide starburst induced by the tidal interaction with the cluster core.  About half of the stars (with significant systematic variance) are metal-poor and have old and intermediate ages, possibly coming from the outer arms of the disrupted spiral. The spectrum also shows a weak metal poor component at $\sim 1$Gyr, which may or not be real.  If real, it may be a component of the ICL rather than of the galaxy itself since the sky subtraction for the arc spectra was done far away from the ICL and the BCG halo (cf. Figure~\ref{fig:slits}).  
 
We remark that our best models for the Arc are not able to fit the CN feature at $\sim 3800$\AA.~ We have not found a satisfactory explanation for this failure; We looked at the populations synthesis fits of thousands of  SDSS galaxies  \citep{Cid2011} and we did not find a single one with such a strong CN feature.  We also investigated the possibility that in part the feature comes from a  higher redshift background galaxy. If the CN feature were contaminated by the MgII $h$ and $k$ resonance doublet at 280~nm,  for example, in order to appear at z=0.29, the redshift of the background source would have to be  $z\sim0.75$, and the separation between the components of the doublet about 9.5\AA. Indeed the fit residuals show a clear doublet structure, but with a separation of $\sim40$\AA,  so it is not MgII at z=0.75. An alternative that fits this doublet could be CaII H \& K from a foreground galaxy, but we do not observe the expected lines at other wavelengths.  In the end we opted for masking-out the whole feature in order to avoid introducing systematic errors in the models.

The velocity dispersion of the Arc (272~km/s; Table~1) is high for isolated spiral galaxies, which should not be surprising since we are observing the galaxy being stretched and distorted by the tidal field of the grinding machine, and since we must be picking up a substantial fraction of the spiral rotation as well.
 

 \subsection{Moving away from the BCG}
 
%

%

\begin{figure*}
  \hspace*{-0.3cm} \includegraphics[height=8cm,width=8.5cm]{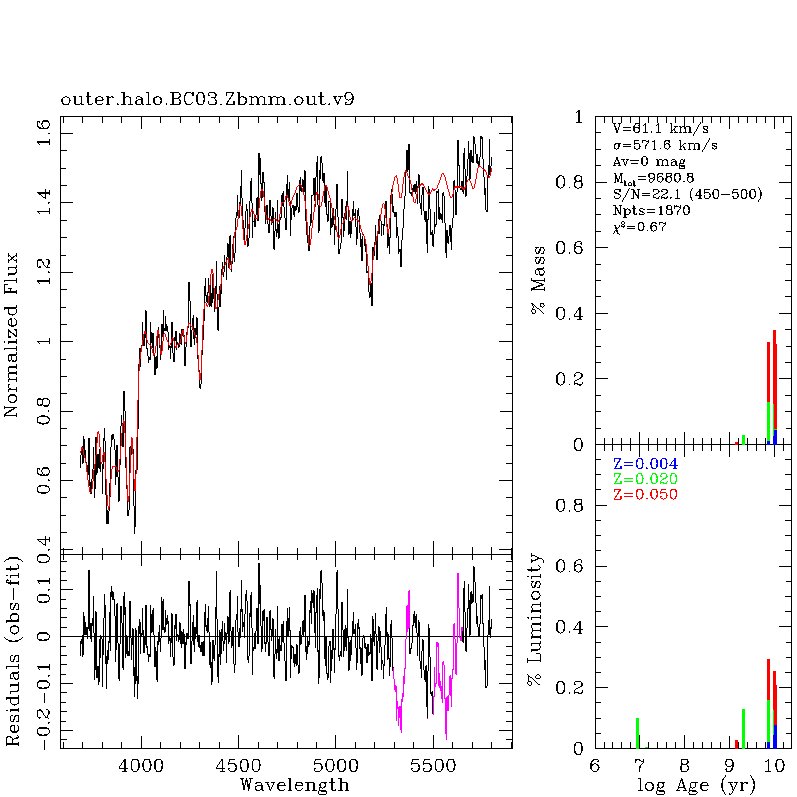}\hspace*{0.5cm}
   \includegraphics[height=8cm,width=8.5cm]{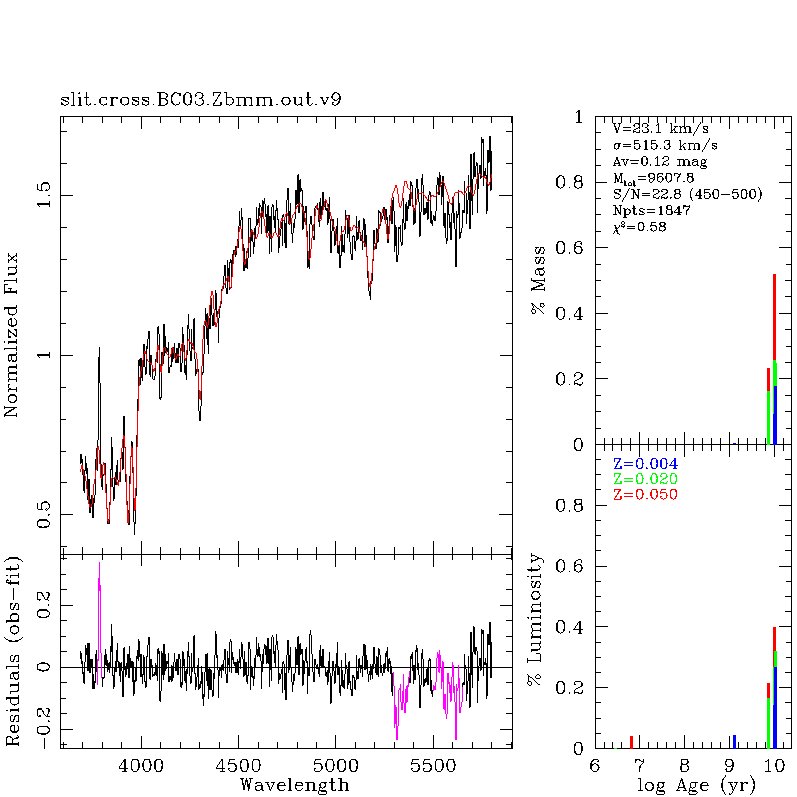}
   \caption {\bf Left. \rm Spectrum of the outer halo of the BCG as indicated by the lower green spot in Figure~\ref{fig:slits}. \bf Right. \rm Spectrum of the position where the southern N-S slit and the E-W slit cross, marked $B$ in Figure~\ref{fig:slits}. The strong unidentified emission-like feature near 3800\AA\ (probably due to an instrumental fluke or improper sky subtraction) was masked from the synthesis as shown in the residuals. }
   \label{fig:mix}
\end{figure*}

We have two more positions along the major axis (as defined by the long axis of the  dumb-bell BCG): the outer halo position indicated by the green spot in the insert of Figure~\ref{fig:slits}, and the crossing of the EW and NS slits also shown in that figure indicated with the letter $B$. The left panel of Figure~\ref{fig:mix} shows the ssp fit for the outer halo. The populations are dominated by old metal-rich stars ($\sim65\%$ of the mass) with a significant fraction $\sim10\%$  of old metal-poor stars.  The solar-abundance components show the largest systematic fluctuations, but the weak population of the same age as the strong burst seen in the S-shaped arc ($\sim2.5$~Gyr) seems to be real.  At these positions we are seeing a mix between ICL and BCG halo, so the solar abundance component is most likely an ICL feature resulting from the disruption of  spiral galaxies that underwent strong starbursts as they approached the core of the cluster, while the metal poor component may arise from the outer parts of these disrupted spiral galaxies, or from disrupted dwarfs.
 
The velocity dispersion at this position (456~km/s; Table~1) is surprisingly high, higher in fact than that of the pure ICL and of the cluster as a whole (see below).  We have done several tests by fitting only parts of the spectrum, but the velocity dispersion remains high. We do not have a good explanation for this high velocity, which we believe to be real; we can only speculate that it may be due to the interaction of the third giant elliptical in the center (\#493 in Figure~\ref{fig:slits}) with the BCG, which could generate high velocity tidal streams. It could also be due to a usual distortion of the halo of the central dumb-bell galaxy, which affects the velocity field in ways that depend on the actual geometry of this subsystem. 

%
            
%

At the location of the slit crossing ($r\sim30$~kpc), the stellar population is completely dominated by old stars of, however, different metallicities. The solar and metal-rich populations suffer the highest systematic variance, while the metal-poor component remains stable. In the standard assumption that the mixed populations are the result of the assembly history of the cluster, we must conclude that the ICL is formed of stars stripped from galaxies of all sizes. However, since most stars have solar and over solar metallicities, the `donors' of ICL were mostly relatively massive galaxies. There are traces (in the luminosity plot) of young stars, but typically these appear when {\bf\tt STARLIGHT} tries to fit low-mass hot post-AGB stars that are not represented in the stellar libraries.

The velocity dispersion of the diffuse component at this distance from the center ($r\simeq40$~kpc) is $\sigma_*\simeq375$~km/s  indicating that this region is mostly dynamically decoupled from the BGC and consists mainly of stars freely moving in the potential well of the cluster.

 
\subsection{The (pure) ICL spectrum}


%
            
%

\begin{figure*}
  \includegraphics[height=11.0cm,width=11.5cm]{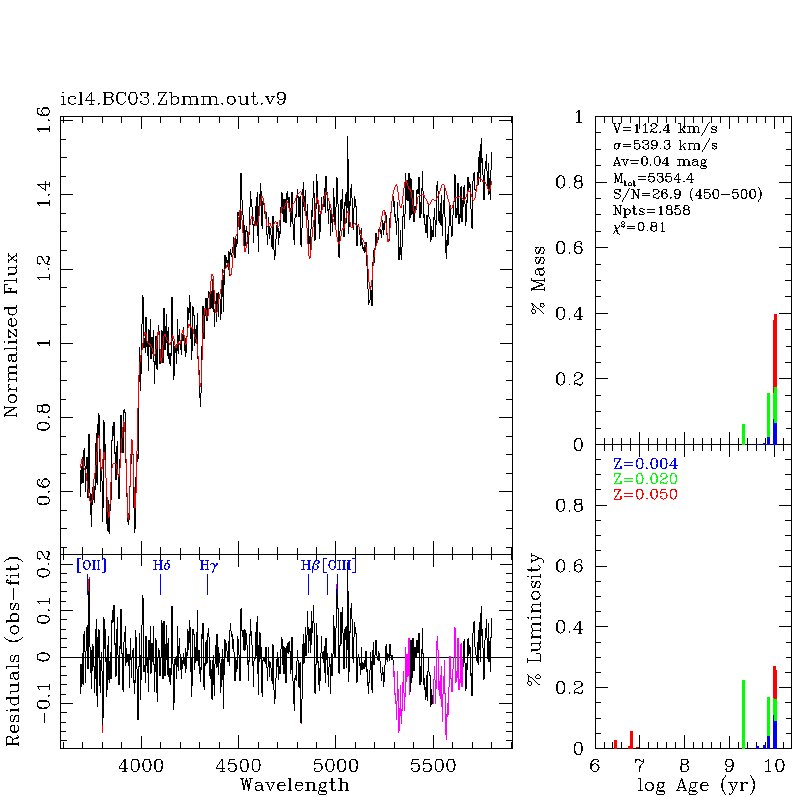}
    \caption{ Stacked spectrum of 4 different regions of the ICL as marked in Figure~\ref{fig:slits}. The subtraction of the synthesis fit allows to unveil faint but clearly visible nebular emission lines of [OII]3727, H$\beta$, and [OIII]5007. Although the residual spectrum is too noisy to justify a detailed analysis of the nebular parameters, the line ratios are at least consistent with those of metal rich HII regions ionized by young massive stars.  There are also hints of the presence of such young massive stars in the population synthesis models. }
   \label{fig:icl}
\end{figure*}

In order to obtain a reasonably high S/N sample of the {\it bona-fide} ICL, which as we saw in Section~\ref{SECicl} corresponds to $r\geq50$~kpc from the BCG, we averaged together 4 regions where the sky-subtracted spectra had reasonably good S/N ratios. These regions are indicated by the red spots in  Figure~\ref{fig:slits}. The resulting combined spectrum shown in Figure~\ref{fig:icl} has $S/N\sim27$ between 450 and 500~nm, which seems rather good considering the low surface brightness levels involved.
 
As expected from the slit-crossing spectrum, the `pure' ICL is dominated by old stars ($91\pm3\%$ of the mass). Of these, $38\pm5\%$  are metal-rich and a similar fraction have solar metallicities (with similar errors). As before, the most stable population is old and metal-poor, whereas the solar components show the largest systematic variances. There are traces of an intermediate-age component of solar metallicity with $\sim8\pm3\%$ of the mass, but again with large systematic variance.
The fraction of luminosity plot (bottom-right) indicate traces of very young stars ($<10^7$) years. These could again be due to hot post-AGB stars, but the residuals plot  (bottom left) shows weak nebular emission lines of [OII]3727, H$\beta$, and [OIII]5007.  It is rather difficult to assess the real strength of H$\beta$ because of the strong underlying absorption feature, but the line is clearly detected. 

The presence of H$\beta$ is important to understand the origin of the nebular lines;  they could arise from the interaction of material ejected from the galaxies with the hot X-ray emitting gas in the cluster, or they could depict in-situ star formation in the ICL as predicted by the cosmological models. The nebular component could also be due to intergalactic planetary nebulae, which are known to be plentiful in other clusters (e.g. \cite{Doherty2009} and references therein).  However, the [OII]/[OIII] line ratios, albeit very noisy, resemble more the spectrum of a metal-rich HII region than that of a PN.


\section{Discussion}
\label{SECres}

Table~\ref{tab:kine} summarizes the kinematical data from the population fits. We see a steep increase in the velocity dispersion with radius qualitatively similar to what is observed by \cite{Coccato2012} in Hydra-I. The 1D radial velocity dispersion of the ICL is 408~km/s with an estimated error of $\pm20$~km/s. The velocity dispersion of the BCG halo is consistent with the values observed in other giant ellipticals and also in the central regions of massive ellipticals. Thus, at that position the light is completely dominated by the BCG, which seems to be at odds with the population synthesis models that show that about 45\% of the halo population is composed of old metal poor stars, whereas for an elliptical galaxy we expect the population to be dominated by metal rich stars (as is the case of Galaxy\#210 for example). This indicates that the BCG halo has a significant component of stars stripped from dwarf galaxies. We have discussed above the large velocity dispersion of the outer halo, for which we do not have a good explanation.

\begin{table*}
\begin{centering}
\caption{Velocity dispersions of the studied components}
\begin{tabular}{cccc}
 \hline
Component  		&	R      & $\sigma_{obs}$ & $\sigma_*$  \\ 
				&	kpc		& km/s	& 	km/s		  \\ \hline
Galaxy \#210		&	---		&  385.2	&	152		\\
BCG Inner Halo  	&	10		&  384.3 	& 	150   \\ 
S-shaped Arc 	  	&	20		&  446.5 	&  	272   \\
BCG Outer Halo 	&	30		&  577.3 	&  	456   \\
Slit Crossing	 	&	40		&  515.9 	&  	375  \\
ICL			 	&	$\geq50$ 	&  540.0 	& 	408 	\\  \hline
 \end{tabular}
 \label{tab:kine}
\end{centering}
\end{table*}

By definition the ICL stars should be freely floating in the gravitational potential of the cluster, and therefore should essentially have a similar velocity dispersion to that of the cluster as a whole. Interestingly, our cluster has a bimodal radial velocity distribution (Paper~I), which yields two different values for the velocity dispersion depending on whether the strongest peak (that contains 90\% of the galaxies) represents the cluster, or the double feature is simply a statistical fluke (we have redshifts for 95 galaxies in the cluster; Paper~I). In the first case we get $\sigma_{cl}=328$ km/s  whereas in the second case, that is including all galaxies, we find  $\sigma_{cl}=496$ km/s.  Our value for the ICL lies exactly between these two extremes, confirming that indeed that at $r\geq50$pc we are seeing the {\it bona-fide} ICL population.

The stellar populations of the ICL are more of a puzzle; The state-of-the-art cosmological simulations of \cite{Puchwein2010} provide detailed predictions for the star formation history of the ICL for a model cluster of $10^{14}h^{-1} \rm M_{\odot}$ at $z=0$, which is reasonably close to the mass of our cluster ($7\times10^{13}\rm M_{\odot}$; Paper~I).  In particular, their Figure~9 plots the distribution of the number of star particles formed as a function of redshift that can be directly compared to the results of our population synthesis models.   The simulations (with AGN feedback) predict that about 25\% of the ICL stars formed more recently than $z\sim1.5$. In other words,  close to 25\% of the ICL stars in our cluster should be younger than about 6~Gyr.  Our observations show that $8\pm3\%$ of the stars were formed at $t \lesssim 6$~Gyr, most of which have solar abundance. Again the systematic variance of this fraction is rather high, but systematic fluctuations never exceed the fraction of 8\% for these `young' stars.  Moreover,  the cosmological simulations predict that close to 30\% of all stars in the ICL are in fact born {\it in situ}, in small metal-poor gas clouds distributed throughout the clusters. These simulations also predict that the vast majority of these clouds fell into the cluster very recently.  Indeed we seem to see traces of a hot, young  ($t<10$~Myr) stellar population, but, if real, this population accounts for significantly less that 1\% of the total stellar mass in the ICL.

It is interesting to compare our results with the populations of the ICL in the Virgo cluster revealed by HST/ACS photometry of resolved stars \citep{Williams2007} and by spectroscopy in Hydra-I  \citep{Coccato2012}.  The Virgo photometry is somewhat difficult to interpret because it depends on what isochrones they use, and in the way  they treat AGB stars. Nevertheless, \cite{Williams2007} state that in their best fitting models the stellar populations are dominated by old metal poor stars, which constitute 70\% to 80\% of the total ICL population. There is also evidence in their photometry for a younger metal-rich component comprising between 20\% and 30\% of the mass.  

Using Lick indices to interpret their data,  \cite{Coccato2012}  find that essentially 100\% of the ICL stars in Hydra-I are old and metal poor.  In contrast,  the ICL in our cluster is dominated by old  {\it metal rich} stars that together with those of solar abundance comprise $75\pm7\%$ of the ICL mass.  Thus, just like in Virgo and Hydra-I, the ICL population of our cluster is dominated by old stars, but in sharp contrast with these clusters,  less than 20\% of the ICL stars in our cluster are metal poor.  
 
 It is tempting to speculate that the intermediate age stars of solar abundance are the remnants of E+A galaxies where the A component is the consequence of a tidally induced, galaxy-wide starburst. In fact, the S-shaped arc in the center of the cluster appears to be  the remnant of one or two spiral galaxies caught in the act of being triturated by the {\it grinding machine} (Paper~I). Our extensive study of the cluster was in fact originally aimed at understanding the nature of the Arc, which in \cite{Faure2007} we ascribed to the gravitational lensing of a pair of background galaxies.  But Figure~\ref{fig:halo} shows that the spectrum of the Arc actually resembles an E+A (`post-starburst') galaxy at the redshift of the cluster, and where $\sim 50\%$ of the stars formed in a galaxy-wide starburst 2.5-4 Gyr ago.
We stress, however,  that we do not fully understand the spectrum of the S-shaped arc as discussed above.

 \section{Conclusions and Future work}
 \label{SECconcl}

By stacking our deep spectra of the ICL of an X-ray cluster at $z=0.29$ we have been able to obtain a combined  spectrum with sufficiently high $S/N$ ratio to decompose the stellar populations of the ICL. We find that the resulting star formation history for the ICL, as told by the population synthesis,  is not consistent with the predictions of state-of-the-art cosmological simulations.  While the overall populations are dominated by old  stars, consistent with the model predictions that the majority of the ICL stars come from massive galaxies, there is a significant fraction of ICL stars that are old and metal poor, indicating that dwarf galaxies, and perhaps the outer regions of spiral galaxies, also contributed significantly to the ICL. Given the structure of the cluster, much of the intergalactic stars probably came from the interaction of the three massive elliptical galaxies at the center, two of which are in an advanced stage of merging revealed by their dumb-bell structure.

We also see an intermediate age population of solar abundance, which could be due to stars stripped off E+A galaxies. This would be the case if spiral galaxies that approach the cluster center undergo galaxy-wide tidally induced starbursts before being disrupted by the {\it grinding machine}.  The population synthesis models also allow us to reveal faint nebular emission lines in the spectrum of the ICL. This  suggest the existence of intergalactic HII regions, which could be ionized by young massive stars  formed in-situ. However, the total mass fraction of these stars is $\ll1\%$ whereas the simulations predict that up to 30\% of the stars would form in the ICL itself. 

It is also possible that the nebular lines arise from unresolved planetary nebulae, although the [OII]/[OIII] ratio does not resemble those of planetaries, but the residual spectrum is too noisy to affirm this with any certainty. Finally, the nebular emission could be due to the interaction of cold gas ejected from the galaxies with the hot X-ray emitting plasma that pervades the cluster (e.g. \cite{Ferland2009}).    

Our observations are in sharp contrast with what is seen in nearby clusters such as Virgo or Hydra-I where the vast majority of the ICL stars are old and metal poor. The fact that our cluster is at a higher redshift ($z\sim0.29$) makes no difference since galaxy clusters are not observed to evolve significantly between $z\sim0.3$ and $z=0$. It seems more reasonable therefore to ascribe the discrepancy to the peculiar structure of our cluster for which the vast majority of the ICL stars were stripped relatively recently after a major merger that brought the 3 giant elliptical galaxies to the cluster center.

It would be of considerable interest to observe a sample of clusters at intermediate   redshifts ($z\sim0.15$ to $z\sim0.25$) to compare the composition of their ICL for different configurations of the close environment of the brightest cluster galaxy.  Observing intermediate redshift clusters allows us to cover a substantial fraction of the ICL within the spectrograph slits. Using MOS instruments would have the advantage of allowing to measure the velocity dispersion of the cluster at the same time as we measure the ICL, as we did for our $z=0.29$ cluster.
 
\section*{Acknowledgements}

Roberto Cid-Fernandez, the father of {\bf\tt STARLIGHT}, kindly provided us with advice about installing and using his code, as well as about the properties of various stellar libraries, and criteria to interpret the model fits and to choose the most appropriate library for our work. This work would not have been possible without Roberto's generous help.

\bsp
\label{lastpage}

\end{document}